\documentclass[prb,aps,twocolumn]{revtex4-1}

\begin{document}
\title{Comment on ``Time-dependent current-density functional theory for
generalized open quantum systems" by
J. Yuen-Zhou, C. Rodr\'iguez-Rosario and A. Aspuru-Guzik, 
{\it Phys. Chem. Chem. Phys.} 2009, 11, 4509}
\author{Roberto D'Agosta}
\affiliation{Nano-Bio Spectroscopy group and ETSF Scientific 
Development Centre,
Dpto F\'isica de Materiales, Universidad del Pa\'is Vasco UPV/EHU, 
E-20018 San Sebasti\'an, Spain}
\affiliation{IKERBASQUE, Basque Foundation for Science, 48011, Bilbao, 
Spain}
\author{
Massimiliano Di Ventra}
\affiliation{Department of Physics, University of California
- San Diego, La Jolla, California, 92093, USA}
\date{August 25, 2009}
\maketitle

In a recent article\cite{Yuen-Zhou2009a} Yuen-Zhuo {\it et al.} (YZ)
have proposed an extension of our theorem of Stochastic
Time-Dependent Current Density Functional Theory
(STDCDFT).\cite{DiVentra2007, DAgosta2008a} The main claim of this
generalization is a mapping between open and closed quantum systems,
namely that the dynamics of the current and particle densities of
an open quantum many-body system can be uniquely determined by the
time evolution of the same quantities of
an effective non-interacting closed system. If
this were true, there would be no need to consider the evolution of
a density matrix via a master equation, but rather the dynamics of
the current and particle densities could be obtained via the time
evolution of the standard Schr\"odinger equation. From a
computational point of view this would be dramatically cheaper than
solving for a master equation,\cite{Pershin2008} and would also
leave open the possibility of analytical solutions, asymptotic
expansions, etc. In this comment we point out some serious
deficiencies in the proof of YZ's theorem that invalidate
their main results.

Let us begin with a critical examination of Theorem 1 in
Ref.~\onlinecite{Yuen-Zhou2009a}. To do this, let us recall that in
the theory of open quantum systems, the equation of motion for the
ensemble-averaged particle density, $n(r,t)$, is given
by\cite{Frensley1990}
\begin{equation}
\partial_t n(r,t)=-\vec{\nabla}\cdot \vec{j}(r,t)+\mathcal{F}_B(r,t)
\end{equation}
where $\mathcal{F}_{B}(r,t)$ describes the density modulation
induced by the presence of the bath and $\vec{j}(r,t)$ is the
ensemble-averaged current density. \footnote{In this theory we do
not allow for the system to exchange particles with the
environment.} Eqn 1 can be obtained from the master equation for the
density matrix, $\hat\rho$
\begin{equation}
\partial_t \hat \rho(t)=-i\left[ \hat H(t),\hat \rho(t)\right] 
+\int_{t_0}^t dt' K(t,t')\hat \rho(t').
\label{master}
\end{equation}
Here, $K(t,t')$ is a memory kernel that describes the action of the
bath on the system, and $\hat H(t)$ is the Hamiltonian of the system
that evolves under the action of an external vector potential
possibly time dependent (we work, as in YZ's work in a gauge in
which the scalar potential has been set to zero).

Theorem 1 in Ref.~\onlinecite{Yuen-Zhou2009a} then states that there
is a one-to-one correspondence between a vector potential and the
pair of functions given by the current and particle densities.
However, this statement cannot be correct as a simple mathematical
characterization of the functional spaces connected by this mapping
shows. If, as YZ seem to argue, the particle and current densities
are {\it independent} functions, then they define a 4-dimensional
functional space. This amounts to saying that $\mathcal{F}_B(r,t)$ is
not completely determined by the sole knowledge of $n(r,t)$ and
$\vec{j}(r,t)$. Since a general vector potential is a vector of
three independent functions, Theorem 1 in YZ's work would imply that
a 4-dimensional functional space formed by the density plus the
three components of the current density is locally homeomorphic to a
3-dimensional functional space spanned by the vector potential. This
is obviously incorrect. A similar problem is present in the
connection between the scalar potential and the current
density.\cite{DAgosta2005a} The solution to this inconsistency is
that the continuity eqn (1) is indeed a non-linear equation in the
particle and current densities from which the particle density can
be obtained starting from the current density, once the bath
operator and initial conditions have been fixed.\cite{DiVentra2007,
DAgosta2008a} This amounts to saying that $\mathcal{F}_B(r,t)$ is a
functional of $n(r,t)$ and $\vec{j}(r,t)$, or better of
$\vec{j}(r,t)$ alone, and that equation (1) admits a unique
physical solution.

It is also worth pointing out that our proof of the
theorem,\cite{DiVentra2007} as well as that of the theorem of
standard TD-CDFT,\cite{Vignale2004} does not require that the
particle density in the auxiliary system is equal to the particle
density in the original one.\footnote{Indeed, in the standard
TD-CDFT proof one uses the standard continuity equation to infer the
equality between the two densities. In our proof, we postulated it
from the uniqueness of the solution of the non-linear continuity eqn
(1), with $\mathcal{F}_B(r,t)$ a functional of the current density.}
In those proofs one only needs to determine the n-th time derivative
of all quantities, and only the (n+1)-th time derivative of the
vector potential. The n-th time derivative of the particle density
is obtained from eqn (1) (or, in a closed system from the same
equation with $\mathcal{F}_B=0$). That equation, however, does not
contain any (n+1)-th time derivative. Therefore, the equation for
the vector potential in the auxiliary system we use in our
proof\cite{DiVentra2007} is still a recursive relation, with a
unique solution provided the initial conditions. The theorem then
guarantees the one-to-one correspondence between external vector
potential and ensemble-averaged current density, leaving quite open
the (possibly difficult) task of obtaining the ensemble-averaged
density from the current density, when the continuity equation is
given by eqn (1).

Apparently, the motivation of YZ's work to extend our STDCDFT
theorem appears connected to this point, namely that eqn~(1) may not
always be satisfied, i.e., that the contribution ${\cal F}_B(r,t)$
is not equal in the real and KS systems, if one keeps the bath
operator fixed [see discussion after eqn~(18) in their manuscript].
Besides the fact that this statement is not mathematically proven -
and following the discussion of the previous paragraph, not strictly
necessary - we could argue that, since in calculating the particle
and current densities one fixes only a few degrees of freedom of the
density matrix, two density matrices can indeed produce the same
average particle and current densities (in two different systems)
and satisfy eqn~(1) with the same bath operator. In addition, unlike
what YZ claim, we have never argued that the term ${\cal F}_B(r,t)$
needs to be small~\footnote{However, in our theory, this term is
proportional to the modulus square of the coupling constant between
the bath and the system, which by assumption is
small.\cite{Gaspard1999} In the formalism proposed by
Zwanzig,\cite{Zwanzig1960} that YZ seem to follow, the degrees of
freedom that are integrated out are not in equilibrium and do not
describe a thermal bath.} or to vanish identically. Indeed, when
discussing the continuity equation (1) in our formalism (sec. II-D
of Ref.~\onlinecite{DAgosta2008a}) we have requested that it
uniquely fixes the ensemble-average particle density once the
average current density is given, with or without the contribution
${\cal F}_B(r,t)$. This is definitely true for the initial
conditions, when the system and bath are (by hypothesis)
uncorrelated and thus ${\cal F}_B(r,t_0)=0$. In our original paper
we have then separated the following discussion in two cases. On the
one hand we have considered that this extra term identically
vanishes at any instant of time. This happens, for example, when we
consider local bath operators. In this case we do not need to go any
further because the density can be obtained uniquely from the
current density in the standard way. On the other hand, if the extra
term ${\cal F}_B(r,t)$ does not vanish identically, the only
physical option left is to assume that eqn (1) uniquely determines
$n(r,t)$ given $\vec{j}(r,t)$, once the initial conditions and bath
operator are fixed. Indeed, in our proof we construct the average
density and current density in time, once the initial conditions are
assigned, and show that the term ${\cal F}_B(r,t)$ is some
functional of the current and particle densities. At this point, the
(now possibly nonlinear) equation of motion (1) for the average
particle density admits one and only one physical solution given the
average current density, once the initial state has been
assigned.\footnote{Here, we cannot exclude the existence of
pathological, non-physical cases for which eqn (1) admits more than
one solution for a given average current and initial condition.}

We now want to critically examine Theorem 3 in
Ref.~\onlinecite{Yuen-Zhou2009a}, where the main result of that paper
is presented: YZ claim to show that it is possible to construct
a unique closed non-interacting quantum system that mimics the
dynamics of $n(r,t)$ and $\vec{j}(r,t)$ of the real open interacting
system. We want to point out that, following YZ's reasoning, one can
in fact find many (possibly infinite) closed non-interacting quantum
systems that reproduce the dynamics of the exact current and
particle densities. However, this is in contrast with Theorem 1,
that claims that given the densities only one such system should
exist. In their proof, at step 1, YZ state that the vector
$\vec{C}(r,t)$ can be uniquely determined by their eqn (23)
\begin{equation}
\vec{C}(r,t)=-\frac{m}{e n(r,t)} \int d^3r \left(\frac{\partial
n(r,t)}{\partial t}+\vec{\nabla}\cdot \vec{j}(r,t)\right)
\end{equation}
once a boundary condition in space that fixes an arbitrary function
of time is assigned. However, the above equation is not the unique
solution to eqn (22) of Ref.~\onlinecite{Yuen-Zhou2009a}, which reads
\begin{equation}
\frac{\partial n(r,t)}{\partial t}=
-\vec{\nabla}\cdot \vec{j}(r,t) - \vec{\nabla}
\cdot \left(\frac{e\vec{C}(r,t)}{m} n(r,t)\right).
\end{equation}
For instance, easily fulfilling the assigned boundary condition, one
can add the curl of an arbitrary vector, $\vec{g}(r,t)$, to the
function $\vec{C}(r,t)$ and still satisfy their eqn (22). Indeed it
is easily proven that $\vec{C}'(r,t)=\vec{C}(r,t)
+(\vec{\nabla}\times\vec{g}(r,t))/n(r,t)$ still satisfies eqn (22)
if $\vec{C}(r,t)$ satisfies the same equation. If we now continue
with the proof and use $\vec{C}'(r,t)$ we arrive at a new vector 
potential
$\vec{A}_{KS}'(r,t)$ that gives the same $n(r,t)$ and $\vec{j}(r,t)$ as
the couple $\vec{A}_{KS}(r,t)$ and $\vec{C}(r,t)$. This, however, is in
contradiction with Theorem 1 that claims a one-to-one mapping
between $n(r,t)$, $\vec{j}(r,t)$ and $\vec{A}_{KS}(r,t)$, when all the
other operators and the boundary condition, are kept
fixed.\footnote{
The effective vector potential entering in the KS
Hamiltonian is ${\vec A}_{KS}+{\vec C}$. Furthemore, we do expect that
${\vec A}_{KS}+{\vec C}\not = {\vec A'}_{KS}+{\vec C'}$. Thus we have
two KS systems producing the same current and particle densities in
contrast with YZ's Theorem 1.} 
This ambiguity reflects the fact that YZ
are trying to mimic the effect of a scalar function - the term
${\cal F}_B(r,t)$ in eqn (1) - with a vector function, $\vec{C}(r,t)$, 
without imposing strict boundary conditions (BCs) on
$\vec{C}(r,t)$.

A simple solution to the aforementioned problem may appear by
setting $\vec{\nabla} \times \vec{C}(r,t)\equiv 0$. However, it is
important to realize that the imposition of certain boundary
conditions on the dynamics of these quantities has a direct impact
on the uniqueness of the results. For example, assuming that
$\vec{C}(r,t)$ reaches a certain uniform limit when $|r|\to \infty$,
might be inconsistent with the bath operator acting on the true
many-body system. Indeed, certain bath operators can be strongly
non-local in space, effectively transferring charge from one region
of space to another, with the two arbitrarily far from each other.
Therefore, the BCs on $\vec{\nabla} \times
\vec{C}(r,t)$ have not a clear physical origin or relation to any
physical observable. Fixing their value to obtain one solution
appears utterly arbitrary. Notice that a similar problem appears
also in the case of the standard theorem of TDDFT (see for example
Ref.~\onlinecite{vanLeeuwen1999}) where the additional boundary
condition $n(r)\vec{\nabla} \Delta V(r,t)\to 0$ when $|r|\to \infty$
is added to the proof. However, in this case, while in principle
this condition is arbitrary and one may choose another condition,
this choice is motivated by physical arguments that are valid for a
wide range of systems. The same considerations instead do not apply
to all the components of the vector $\vec{C}(r,t)$. Therefore, the
proof of the theorem as it is formulated in
Ref.~\onlinecite{Yuen-Zhou2009a} cannot hold for general bath
operators. 

Finally, we want to comment on a fundamental but important issue.
YZ's initial assumption for their theorem is a closed equation of
motion for the density matrix. As we have discussed at length in our
previous publications,\cite{DiVentra2007, DAgosta2008a} this is not
a solid starting point for a formulation of DFT for open quantum
systems. This is due to both the possible loss of positivity of the
density matrix if an equation of motion of such quantity is employed
with the Hamiltonian and/or bath operator(s) dependent on
time,\cite{Ford2005} and the fact that the KS Hamiltonian does
depend on internal degrees of freedom. Starting from the master
equation formulation of the same problem, one needs to exclude from
the outset the possibility that the Hamiltonian of any auxiliary
system with different interaction potential (and hence the KS
Hamiltonian) depends on the internal degrees of freedom. Otherwise,
for such a system no closed density-matrix equation can be obtained.
In other words, one needs to start from an hypothesis that
constitutes part of the final thesis. It is only when one starts
from a stochastic Schr\"odinger equation for the {\em state vector}
that one can prove that the {\em exact} KS Hamiltonian depends only
on the average current density.

RD acknowledges the hospitality of Imperial College London
where part of this work has been completed.
MD acknowledges financial support from the Department of Energy grant 
DE-FG02-05ER46204.

\bibliographystyle{rsc}
\bibliography{articles,mine,books}

\providecommand*{\mcitethebibliography}{\thebibliography}
\csname @ifundefined\endcsname{endmcitethebibliography}
{\let\endmcitethebibliography\endthebibliography}{}
\begin{mcitethebibliography}{16}
\providecommand*{\natexlab}[1]{#1}
\providecommand*{\mciteSetBstSublistMode}[1]{}
\providecommand*{\mciteSetBstMaxWidthForm}[2]{}
\providecommand*{\mciteBstWouldAddEndPuncttrue}
  {\def\EndOfBibitem{\unskip.}}
\providecommand*{\mciteBstWouldAddEndPunctfalse}
  {\let\EndOfBibitem\relax}
\providecommand*{\mciteSetBstMidEndSepPunct}[3]{}
\providecommand*{\mciteSetBstSublistLabelBeginEnd}[3]{}
\providecommand*{\EndOfBibitem}{}
\mciteSetBstSublistMode{f}
\mciteSetBstMaxWidthForm{subitem}
{(\emph{\alph{mcitesubitemcount}})}
\mciteSetBstSublistLabelBeginEnd{\mcitemaxwidthsubitemform\space}
{\relax}{\relax}

\bibitem[Yuen-Zhou \emph{et~al.}(2009)Yuen-Zhou, Rodr\'iguez-Rosario, and
  Aspuru-Guzik]{Yuen-Zhou2009a}
J.~Yuen-Zhou, C.~Rodr\'iguez-Rosario and A.~Aspuru-Guzik, \emph{Phys. Chem.
  Chem. Phys.}, 2009, \textbf{11}, 4509\relax
\mciteBstWouldAddEndPuncttrue
\mciteSetBstMidEndSepPunct{\mcitedefaultmidpunct}
{\mcitedefaultendpunct}{\mcitedefaultseppunct}\relax
\EndOfBibitem
\bibitem[Di~Ventra and D'Agosta(2007)]{DiVentra2007}
M.~Di~Ventra and R.~D'Agosta, \emph{Phys. Rev. Lett.}, 2007, \textbf{98},
  226403\relax
\mciteBstWouldAddEndPuncttrue
\mciteSetBstMidEndSepPunct{\mcitedefaultmidpunct}
{\mcitedefaultendpunct}{\mcitedefaultseppunct}\relax
\EndOfBibitem
\bibitem[D'Agosta and Di~Ventra(2008)]{DAgosta2008a}
R.~D'Agosta and M.~Di~Ventra, \emph{Phys. Rev. B}, 2008, \textbf{78},
  165105\relax
\mciteBstWouldAddEndPuncttrue
\mciteSetBstMidEndSepPunct{\mcitedefaultmidpunct}
{\mcitedefaultendpunct}{\mcitedefaultseppunct}\relax
\EndOfBibitem
\bibitem[Pershin \emph{et~al.}(2008)Pershin, Dubi, and Di~Ventra]{Pershin2008}
Y.~V. Pershin, Y.~Dubi and M.~Di~Ventra, \emph{Phys. Rev. B}, 2008,
  \textbf{78}, 054302\relax
\mciteBstWouldAddEndPuncttrue
\mciteSetBstMidEndSepPunct{\mcitedefaultmidpunct}
{\mcitedefaultendpunct}{\mcitedefaultseppunct}\relax
\EndOfBibitem
\bibitem[Frensley(1990)]{Frensley1990}
W.~R. Frensley, \emph{Rev. Mod. Phys.}, 1990, \textbf{62}, 745\relax
\mciteBstWouldAddEndPuncttrue
\mciteSetBstMidEndSepPunct{\mcitedefaultmidpunct}
{\mcitedefaultendpunct}{\mcitedefaultseppunct}\relax
\EndOfBibitem
\bibitem[Not()]{Note1}
In this theory we do not allow for the system to exchange particles with the
  environment.\relax
\mciteBstWouldAddEndPunctfalse
\mciteSetBstMidEndSepPunct{\mcitedefaultmidpunct}
{}{\mcitedefaultseppunct}\relax
\EndOfBibitem
\bibitem[D'Agosta and Vignale(2005)]{DAgosta2005a}
R.~D'Agosta and G.~Vignale, \emph{Phys. Rev. B}, 2005, \textbf{71},
  245103\relax
\mciteBstWouldAddEndPuncttrue
\mciteSetBstMidEndSepPunct{\mcitedefaultmidpunct}
{\mcitedefaultendpunct}{\mcitedefaultseppunct}\relax
\EndOfBibitem
\bibitem[Vignale(2004)]{Vignale2004}
G.~Vignale, \emph{Phys. Rev. B}, 2004, \textbf{70}, 201102(R)\relax
\mciteBstWouldAddEndPuncttrue
\mciteSetBstMidEndSepPunct{\mcitedefaultmidpunct}
{\mcitedefaultendpunct}{\mcitedefaultseppunct}\relax
\EndOfBibitem
\bibitem[Not()]{Note2}
Indeed, in the standard TD-CDFT proof one uses the standard continuity equation
  to infer the equality between the two densities. In our proof, we postulated
  it from the uniqueness of the solution of the non-linear continuity eqn (1),
  with $\protect \mathcal {F}_B(r,t)$ a functional of the current
  density.\relax
\mciteBstWouldAddEndPunctfalse
\mciteSetBstMidEndSepPunct{\mcitedefaultmidpunct}
{}{\mcitedefaultseppunct}\relax
\EndOfBibitem
\bibitem[Not()]{Note3}
However, in our theory, this term is proportional to the modulus square of the
  coupling constant between the bath and the system, which by assumption is
  small.\cite {Gaspard1999} In the formalism proposed by Zwanzig,\cite
  {Zwanzig1960} that YZ seem to follow, the degrees of freedom that are
  integrated out are not in equilibrium and do not describe a thermal
  bath.\relax
\mciteBstWouldAddEndPunctfalse
\mciteSetBstMidEndSepPunct{\mcitedefaultmidpunct}
{}{\mcitedefaultseppunct}\relax
\EndOfBibitem
\bibitem[Not()]{Note4}
Here, we cannot exclude the existence of pathological, non-physical cases for
  which eqn (1) admits more than one solution for a given average current and
  initial condition.\relax
\mciteBstWouldAddEndPunctfalse
\mciteSetBstMidEndSepPunct{\mcitedefaultmidpunct}
{}{\mcitedefaultseppunct}\relax
\EndOfBibitem
\bibitem[Not()]{Note5}
The effective vector potential entering in the KS Hamiltonian is ${\mathaccent
  "017E\relax A}_{KS}+{\mathaccent "017E\relax C}$. Furthemore, we do expect
  that ${\mathaccent "017E\relax A}_{KS}+{\mathaccent "017E\relax C}\not =
  {\mathaccent "017E\relax A'}_{KS}+{\mathaccent "017E\relax C'}$. Thus we have
  two KS systems producing the same current and particle densities in contrast
  with YZ's Theorem 1.\relax
\mciteBstWouldAddEndPunctfalse
\mciteSetBstMidEndSepPunct{\mcitedefaultmidpunct}
{}{\mcitedefaultseppunct}\relax
\EndOfBibitem
\bibitem[van Leeuwen(1999)]{vanLeeuwen1999}
R.~van Leeuwen, \emph{Phys. Rev. Lett.}, 1999, \textbf{82}, 3863\relax
\mciteBstWouldAddEndPuncttrue
\mciteSetBstMidEndSepPunct{\mcitedefaultmidpunct}
{\mcitedefaultendpunct}{\mcitedefaultseppunct}\relax
\EndOfBibitem
\bibitem[Ford and O'Connell(2005)]{Ford2005}
G.~W. Ford and R.~F. O'Connell, \emph{Ann. Phys.}, 2005, \textbf{319},
  348\relax
\mciteBstWouldAddEndPuncttrue
\mciteSetBstMidEndSepPunct{\mcitedefaultmidpunct}
{\mcitedefaultendpunct}{\mcitedefaultseppunct}\relax
\EndOfBibitem
\bibitem[Gaspard and Nagaoka(1999)]{Gaspard1999}
P.~Gaspard and M.~Nagaoka, \emph{J. Chem. Phys.}, 1999, \textbf{111},
  5676\relax
\mciteBstWouldAddEndPuncttrue
\mciteSetBstMidEndSepPunct{\mcitedefaultmidpunct}
{\mcitedefaultendpunct}{\mcitedefaultseppunct}\relax
\EndOfBibitem
\bibitem[Zwanzig(1960)]{Zwanzig1960}
R.~Zwanzig, \emph{J. Chem. Phys.}, 1960, \textbf{33}, 1338\relax
\mciteBstWouldAddEndPuncttrue
\mciteSetBstMidEndSepPunct{\mcitedefaultmidpunct}
{\mcitedefaultendpunct}{\mcitedefaultseppunct}\relax
\EndOfBibitem
\end{mcitethebibliography}
\end{document}